\documentclass[aip,rsi,amsmath,amssymb,reprint]{revtex4-1}
\usepackage{graphicx}
\usepackage[font=footnotesize,centerlast]{caption}

\begin{document}


\title{Magnetised Thermal Self-focusing and Filamentation of Long-Pulse Lasers in Plasmas Relevant to Magnetised ICF Experiments} 

\author{H. C. Watkins}
\email[]{h.watkins15@imperial.ac.uk}
\affiliation{The Blackett Laboratory, Imperial College London, London, United Kingdom}

\author{R. J. Kingham}
\affiliation{The Blackett Laboratory, Imperial College London, London, United Kingdom}


\date{\today}

\begin{abstract}
In this paper we study the influence of the magnetised thermal conductivity on the propagation of a nanosecond $10^{14} \mathrm{Wcm}^{-2}$ laser in an underdense plasma by performing simulations of a paraxial model laser in a plasma with the full Braginskii magnetised transport coefficients. Analytic theory and simulations show the shortening of the self-focal length of a laser beam in a plasma as a result of the reduction of the plasma thermal conductivity in a magnetic field. Furthermore the filamentation of a laser via the thermal mechanism is found to have an increased spatial growth rate in a magnetised plasma. We discuss the effect of these results on recent magnetised inertial fusion experiments where filamentation can be detrimental to laser propagation and uniform laser heating. We conclude the application of external magnetic fields to laser-plasma experiments requires the inclusion of the extended electron transport terms in simulations of laser propagation.
\end{abstract}


\maketitle 

\section{Introduction}

Laser self-focusing in a plasma is the effect in which the laser causes the local plasma refractive index to change by digging a density channel. The beam then refracts into this channel in a process that feeds back and becomes unstable and forms laser filaments \cite{kelley1965self,Kaw1988}. This process can occur via thermal, ponderomotive and relativistic mechanisms and can be detrimental to the performance of Inertial Confinement Fusion (ICF) experiments as the nonuniformity of the beam can seed hydrodynamic instabilities in direct drive\cite{Craxton2015} and the high-intensity laser fields in the filaments can make the experiment more susceptible to parametric instabilities \cite{Baldis1992ParametricPlasmas}. 

The pursuit of better performance in ICF experiments has led to proposing the use of applied magnetic fields\cite{Perkins2017TheFusion,Chang2011FusionImplosions,Montgomery2015UseLaser-coupling}. The hohlraum of an indirect drive experiment is magnetised by an axial magnetic field from an external Helmholtz coil with the aim to reduce the perpendicular heat flow and localise alpha particle transport; thus relaxing the conditions required for thermonuclear burn. Likewise, the Magnetised Liner Inertial Fusion (MAGLIF) scheme\cite{Hohenberger2012,Slutz2012,Harvey-Thompson2015a} relies on an axial magnetic field to reduce the perpendicular heat flow in the DT fuel within the liner. 

The presence of laser propagation through a magnetised plasma in these experiments leads to the question of the effect of the magnetic field on the laser propagation through the plasma itself.

In laser plasmas the plasma beta, the ratio of thermal to magnetic pressure, is very large and so the magnetic pressure is insignificant. However in magnetised high energy density environments referenced above the dimensionless Hall parameter, defined as the product of the electron cyclotron frequency $\omega_e$ and the electron-ion collision time $\tau_{ei}$
\begin{equation}
\chi = \omega_e \tau_{ei},
\end{equation}

reaches the order of 1-10 and so the transport becomes magnetised. 

Electrons in a magnetic field travel on circular orbits perpendicular to the direction of the magnetic field. As such the electron mean free path is magnetically confined and the collisional process of thermal transport is reduced perpendicular to the magnetic field. Following the formalism of Braginskii\cite{S.I.Braginskii1965} this magnetisation causes thermal transport to deviate from the classical isotropic form,
\begin{eqnarray}
\kappa_{clas}=\frac{128}{3\pi}\zeta(Z) n_e v_{th}\lambda_{ei}.
\end{eqnarray}
This is the product of the electron number density $n_e$, the electron thermal speed $v_{th}$, the electron-ion collision length $\lambda_{ei}$ and the function $\zeta(Z)$ which is
\begin{eqnarray}
\zeta(Z)=\frac{Z+0.24}{Z+4.2}.
\end{eqnarray}
Along with the factor $128/3\pi$, this is taken from Epperlein\cite{Epperlein1991APlasmas} and accounts for an electron - electron collisional correction to the conductivity; henceforth we will use $\kappa^{(1)}=\frac{128}{3\pi}\zeta(Z)$. The thermal conductivity becomes anisotropic to a degree dependent on the Hall parameter. In this formalism the electron heat flow driven by a temperature gradient splits the familiar Fourier law into the parallel ($\kappa_\parallel$), perpendicular ($\kappa_\perp$) and Righi-Leduc ($\kappa_\wedge$) terms relative to the direction of the magnetic field $\textbf{b}$,
\begin{equation}
\textbf{q}=-\kappa_\parallel \textbf{b}(\textbf{b}\cdot\nabla T)-\kappa_\perp \textbf{b}\times(\nabla T \times\textbf{b})-\kappa_\wedge\textbf{b}\times \nabla T.
\end{equation}

The magnetised reduction approximately follows the relation
\begin{equation}
\kappa_\perp \sim \frac{\kappa_\parallel}{1+\chi^2},
\end{equation}
where the thermal conductivity perpendicular to the magnetic field is reduced by a factor that is a function of the Hall parameter.

The functional form of the magnetised perpendicular thermal conductivity we will use is
\begin{equation}
\kappa_\perp = \kappa_{clas}\kappa_\perp^c(\chi,Z).
\end{equation}
The factor $\kappa_\perp^c(\chi,Z)$ corresponds to a normalised Hall parameter dependent correction to account for the magnetisation. In the analysis and the numerical simulations that follow the explicit expression of this term is the Epperlein and Haines form \cite{Epperlein1986PlasmaEquation} obtained via a polynomial fit to a numerical solution of the electron Vlaslov-Fokker-Planck equation. However, unlike in the Epperlein-Haines expression, in this paper we will take it to be the normalised form such that $\kappa_\perp^c(0,Z)=1$.  

The thermal mechanisms of self-focusing and filamentation of a laser are mediated by the thermal conductivity \cite{Epperlein1992NonlocalPlasmas}; as such we expect these processes to be modified by the plasma magnetisation and importantly the factor $\kappa^c_\perp(\chi,Z)$. There has been limited consideration of the influence of magnetized transport effects in existing theoretical and simulation studies of self-focusing and filamentation. 

Read et al.\cite{read2016influence} report simulations using a coupled paraxial laser MHD-Braginskii transport code, highlighting the role of Nernst advection of B-field on self-focusing. Here we develop a theoretical model that elucidates the effect of magnetised thermal conductivity on self-focusing length. We show that this is in qualitatively good agreement with paraxial-MHD-Braginskii simulations. We also consider for the first time the effect of  the magnetic modification of thermal conductivity on filamentation, demonstrating via a theoretical model that applied magnetic fields can enhance the spatial growth rate.

The paper is organised as follows; in Sec. \ref{sec:model} the equation set and starting point for the analytic model is introduced, in Sec. \ref{sec:focus} an analytic macroscopic focusing model is derived and compared to fully nonlinear simulations with a laser-plasma code. In Sec. \ref{sec:fil} the filamentation model is derived and the influence of an applied magnetic field analysed and compared with filamentation simulations. The nonlocal kinetic effects are also discussed. Finally in Sec. \ref{sec:influence} the impact of this magnetised laser propagation is discussed in the context of magnetised laser-fusion experiments.  

\section{Mathematical Model} \label{sec:model}
The analytical starting point for our investigation is a static laser plasma model following from Epperlein \cite{Epperlein1990KineticPlasmas}, Kaw \cite{Kaw1988} and Schmitt \cite{Schmitt1988}. However in contrast we account for the tensor thermal conductivity in the Braginskii formalism of a magnetised plasma\cite{S.I.Braginskii1965}.

This model consists of the reduced electron momentum equation, energy equation and the paraxial Helmholtz equation for the propagation of the laser.
\begin{eqnarray}
& n \nabla_\perp T + T \nabla_\perp n= -\frac{1}{2} \varepsilon_0 \frac{n}{n_c}\nabla_\perp |\psi|^2, \\
&-\nabla_\perp \cdot \left(\underline{\underline{\kappa}}\cdot\nabla_\perp T \right)=\frac{1}{2} \varepsilon_0 \frac{n}{n_c}\nu_{ei} |\psi|^2, \\
&2 i k_0\partial_z \psi +\nabla_\perp^2 \psi+k^2_0 \delta \epsilon\psi=0.
\end{eqnarray}
The variables present are the electron number density $n$ the electron temperature $T$ and the electric field envelope of the laser $\psi$. The laser wavevector is directed along the z direction with wavenumber $k_0$ which also defines the critical density of the plasma $n_c$. The factor $\varepsilon_0$ is the vacuum permittivity. In this paraxial laser model we assume the variation in $n$ and $T$ along the beam is much smaller than across (that is perpendicular $\perp$ to the laser wavevector) the beam, allowing us to ignore the parallel derivatives. 

The plasma is coupled to the paraxial equation via the dielectric variation $\delta\epsilon$ which in this paper is defined as 
\begin{equation}
\delta\epsilon = -\frac{\delta n}{n_c}.
\end{equation}
$\delta n$ is the difference between the `background' electron density where there is no laser field and the local electron number density.
The laser envelope couples to the plasma in the momentum equation via the ponderomotive force, representing a balance between the ponderomotive force and the electron pressure. In the energy equation the laser couples through an inverse bremsstrahlung heating source term which is proportional to the electron-ion collision frequency $\nu_{ei}$, giving an energy balance between laser heating and thermal dissipation.

\section{Magnetised Thermal Self-focusing}\label{sec:focus}
The development of a semi-analytic model for the self-focusing of the beam will enable the determination of the effect of the thermal conductivity on the propagation of a Gaussian beamlet or envelope. The theory of self-focusing here follows from the approach described by He\cite{he2014nonlinear} for neutral media. The aim is to determine the plasma response to the laser in terms of the density change $\delta n$.

Using the cylindrically symmetric Gaussian beam approximation, the laser envelope $\psi$ is assumed to have the form 
\begin{eqnarray}
&\psi=A(r,z)e^{i k_0 S(r,z)},\\
&|\psi|^2=A^2=\frac{A^2_0}{\alpha(z)^2}e^{-\frac{r^2}{a_0^2 \alpha^2}},\\
&S=\beta (z)\frac{r^2}{2}+\phi (z),
\end{eqnarray}
which satisfy the boundary conditions
\begin{eqnarray}
\begin{aligned}
&\alpha(0)=1,\\
&\phi(0)=0,\\
&\beta(0)=\frac{1}{R},\\
&A^2(r,0)=A^2_0 e^{-\frac{r^2}{a^2_0}}.
\end{aligned}
\end{eqnarray}
This approximation means the beam has an initial beam waist of size $a_0$ positioned at $z=0$ and a radius of curvature of the phase front of $R$. Using Eq. 11 the paraxial equation (Eq. 9) becomes the coupled pair
\begin{eqnarray}
\begin{aligned}
&2\partial_z S +\left(\frac{\partial S}{\partial r}\right)^2=\frac{1}{k^2_0 A}\left(\frac{\partial^2 A}{\partial r^2}+\frac{1}{r}\frac{\partial A}{\partial r}\right) +\delta \epsilon, \\
&\partial_z A^2 +A^2 \left(\frac{\partial^2 S}{\partial r^2}+\frac{1}{r}\frac{\partial S}{\partial r}\right) +\frac{\partial S}{\partial r}\frac{\partial A^2}{\partial r}=0.
\end{aligned}
\end{eqnarray}
The dielectric variation $\delta\epsilon$ can be represented in terms of orders of $r$,
\begin{equation}
\delta\epsilon(z,r)=\delta\epsilon_0(z)+\delta\epsilon_1(z)r+\delta\epsilon_2(z)r^2+O(r^3).
\end{equation}
Using Eqs. 12 and 13 we can transform these equations into the following set by considering each order of $r$,  
\begin{eqnarray}
&\alpha ' =\beta \alpha, \\
&\beta'+\beta^2=\frac{1}{k^2_0 a^4_0 \alpha^4}+\delta\epsilon_2,\\
&2\phi'=-\frac{2}{k^2_0 a^2_0 \alpha^2}+\delta\epsilon_0.
\end{eqnarray}
The primes represent derivatives with respect to $z$. Furthermore, by taking the derivative of Eq. 17 we find
\begin{equation}
\alpha''=\left(\beta'+\beta^2\right)\alpha.
\end{equation}
The focal spot of the beam is thus the point at which the normalised Gaussian variance $\alpha^2$ is a minimum.

\subsection{Thermal Mechanism}
The magnetised thermal effect appears through the heat flow. When the applied magnetic field is parallel with the direction of the laser wavevector, the problem becomes cylindrically symmetric, so the energy balance equation (Eq. 8) becomes
\begin{equation}
-\frac{1}{r}\frac{\partial}{\partial r}\left(r\kappa_\perp\frac{\partial T}{\partial r}\right)=\varepsilon_0\frac{n }{2n_c}\nu_{ei}|\psi|^2.
\end{equation}

This equation and the momentum equation are nonlinear, therefore to find an analytic solution we linearise the momentum and energy equations with small perturbations about a background constant density and temperature
\begin{eqnarray}
\begin{aligned}
& T=T_0 +\delta T,\\
& n=n_0 +\delta n,
\end{aligned}
\end{eqnarray}
with these approximations the energy and momentum equations become
\begin{eqnarray}
\begin{aligned}
&n_0\frac{\partial \delta T}{\partial r}+T_0\frac{\partial \delta n}{\partial r}=-\varepsilon_0\frac{n_0}{2 n_c}\frac{\partial|\psi|^2}{\partial r},\\
&-\frac{1}{r}\frac{\partial}{\partial r}\left(r\kappa^0_\perp \frac{\partial \delta T}{\partial r}\right)=\varepsilon_0\frac{n_0 }{2n_c}\nu^0_{ei}|\psi|^2.
\end{aligned}
\end{eqnarray}
$\kappa^0_\perp$, and $\nu^0_{ei}$ are the constant thermal conductivity and collision frequency, as functions of the background density and temperature respectively.

Eqs. 10,12,18, 20, and 23 are then used to derive an equation for the normalised beam waist $\alpha(z)$. The full derivation can be found in appendix A along with the definition of all parameters. The result of the derivation is \begin{equation}
\alpha^2(z)=\frac{1}{2}(1+\frac{c_1}{c_2})+\frac{1}{2}(1-\frac{c_1}{c_2})\cos(\sqrt{c_2} z).
\end{equation}

This expression gives the variance of the beam along the propagation direction, the focal point $z_f$ of the beam being the $z$ position where the variance is a minimum
\begin{equation}
z_f = \frac{\pi}{2 \sqrt{c_2}},
\end{equation}
which can be written in more intuitive form in terms of the peak laser  intensity $I_0$ and the electron-ion collision length $\lambda_{ei}$
\begin{equation}
z_f = \pi\frac{n_c}{n_0}\lambda_{ei} \sqrt{\frac{c n_0 T_0}{I_0}}\sqrt{\kappa^{(1)}}\sqrt{\kappa_\perp^c}.
\end{equation}

\subsection{Effect of the Magnetic Field on focusing}
The factor $\kappa_\perp^c$ appears in the expression for the focal length, therefore as the thermal conductivity decreases with increasing magnetisation, the focal length will shorten. With a greatly reduced heat flow, gradients in the temperature profile are maintained. This leads to a higher pressure gradient and deeper electron density channel formation. The refractive index changes more and the beam refracts with a shorter focal length.

\begin{figure}
\includegraphics[width=\columnwidth]{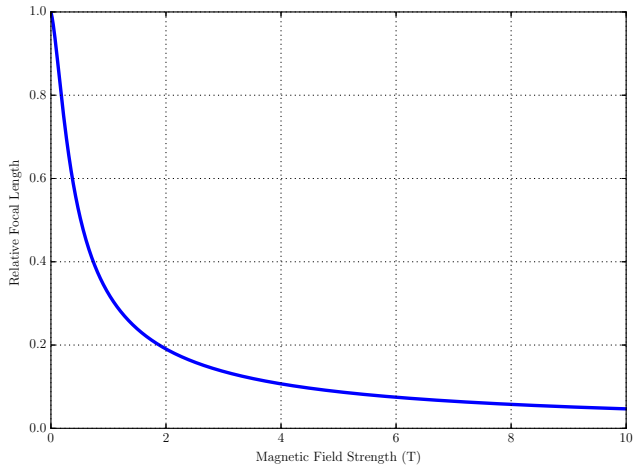}
\caption{The focal length of the beam relative to the unmagnetised case shows the focal length decreases with an increasing magnetic field strength. In a plasma with 1 keV electron temperature and density of $1\times10^{20}\mathrm{cm}^{-3}$ the length drops sharply, at lower temperatures the gradient becomes shallower.  }
\label{fig:1}
\end{figure}

This is illustrated in Fig. \ref{fig:1}. According to Eq. 26, for a $Z=1$, 1 keV temperature, $1\times 10^{20}\mathrm{cm}^{-3}$ number density plasma the focal length of a 1 $\mu$m wavelength, $1\times10^{14} \mathrm{Wcm}^{-2}$ Gaussian beam under a 2 Tesla axial field (magnetising the plasma to a Hall parameter of $\chi=4.8$) is 0.2 of the unmagnetised focal length. In the highly magnetised regime corresponding to $\chi >>1$, the relative focal length is inversely proportional to the applied field strength B.

\begin{figure}
\includegraphics[width=\columnwidth]{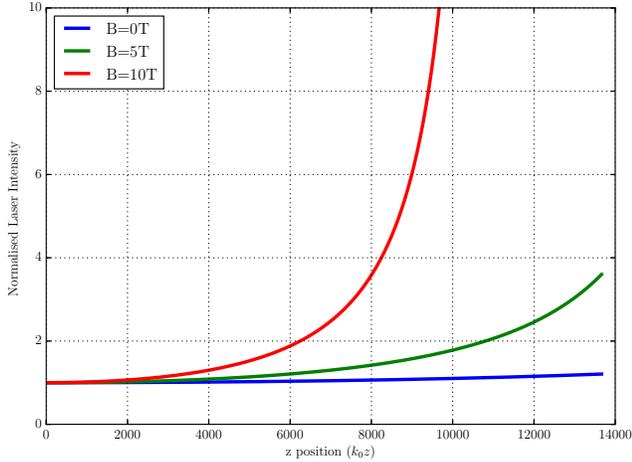}
\caption{The normalised axial laser intensity from the analytic model shows the focusing of the beam when an axial magnetic field is applied with a field strength of 0T, 5T and 10T. The beam focuses over a shorter distance when the magnetic field strength increases. }
\label{fig:2}
\end{figure}

Looking at the intensity of the laser as it undergoes self focusing also shows significant impact of the magnetic field. The axial intensity of a Gaussian beam normalised to the initial peak intensity follows the expression
\begin{equation}
\frac{I(z,r=0)}{I_0}=\frac{1}{\alpha(z)^2}.
\end{equation}
Therefore the axial intensity can be found from Eq. 24 and this is plotted in Fig. \ref{fig:2} for a 0,5 and 10 Tesla applied field with the same parameters as Fig. \ref{fig:1}. Comparing the three curves we find the axial intensity grows faster over a much shorter length to be multiple times the initial intensity as the applied magnetic field increases.  

\subsection{Simulations of the self-focusing of a Gaussian beam in a magnetised plasma}

To compare this theory with the nonlinear regime a simulation was performed with the PARAMAGNET Eulerian magnetised laser-plasma code. The PARAMAGNET code is an implicit single fluid plasma code with the full Braginskii electron transport coefficients using the Epperlein and Haines polynomial fits\cite{Epperlein1986PlasmaEquation}. The plasma is coupled to a paraxial laser model by the ponderomotive force and inverse bremsstrahlung.

A 1 $\mu$m wavelength laser with a cylindrical Gaussian profile and peak intensity of $1\times10^{14} \mathrm{Wcm}^{-2}$ is incident on a pre-ionised $Z=1$, 20 eV plasma with an initially uniform electron density of $1\times 10^{20}\mathrm{cm}^{-3}$. The beam has an initial waist size of 100 $\mu$m at the edge of the simulation domain and propagates through the 3 mm plasma under an applied magnetic field of 0,5 and 10T orientated parallel to the beam wavevector.  

\begin{figure}
\includegraphics[width=\columnwidth]{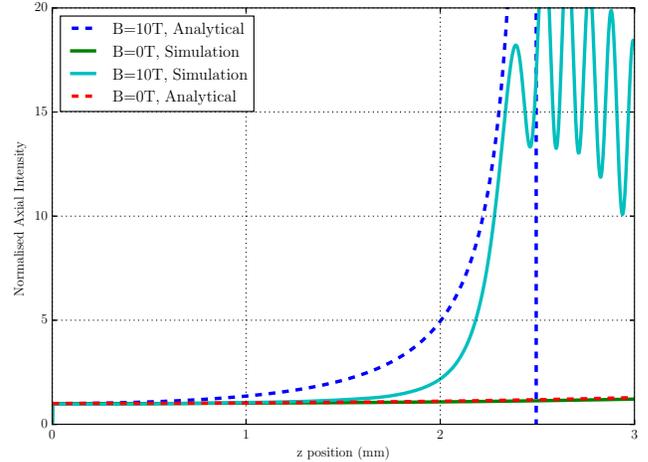}
\caption{The normalised axial laser intensity is compared for the simulated and analytic models for an unmagnetised laser focusing over 3 mm and for a laser-plasma magnetised with a 10T axial field. The analytic model shows good agreement with the simulation up to the first focal point where the analytic model breaks down.}
\label{fig:3}
\end{figure}

Fig. \ref{fig:3} shows the comparison between the axial laser intensity extracted from the simulation after 100ps and the analytic model given by Eq. 27. Two cases are compared, one with a 10T field applied axially with the beam and the other without a magnetic field. In the magnetised simulation the position of the focal point matches closely with the model but the model breaks down when it hits a singularity. At this point the simulation begins to see oscillating behavior where the beam focusses and defocusses in and out of the density channel.

Despite the approximations made in the derivation of Eq. 24 the fully nonlinear PARAMAGNET simulation shows Eq. 24 reproduces the correct focal point position but underestimates the gradient of the intensity increase leading up to the focal point. As with the analytic model the intensity grows significantly over the spatial scale of the simulation to multiple times the initial intensity. The simulation domain was chosen as 3 mm to be similar to laser-underdense plasma experimental conditions such as the preheating stage of MAGLIF\cite{Sefkow2016a}. This implies the laser propagating in such an experiment will focus significantly over the length of several millimeters in a way not present without consideration of the magnetisation of the thermal conductivity.    

\section{Magnetised Thermal Filamentation}\label{sec:fil}
Self-focusing of an optical beam is closely related to filamentation which here will be defined as the breakup of a laser beam seeded by perturbations perpendicular to the direction of beam propagation. This also occurs via a thermal mechanism and will also be influenced by the magnetised thermal conductivity in the same manner.

To derive the spatial growth rate $\gamma$ of the filamentation of a perturbation mode $\textbf{k}_\perp$ perpendicular to the laser with wavenumber $k_0$, we follow the approach taken by Epperlein \cite{Epperlein1990KineticPlasmas,Epperlein1991KineticPlasmas}. In this case the steady-state Eqs. 7-9 are used and linearised with
\begin{eqnarray}
\begin{aligned}
&T=T_0+\delta T e^{(i \textbf{k}_\perp\cdot \textbf{x}-\gamma z)},\\
&n=n_0+\delta n e^{(i \textbf{k}_\perp\cdot \textbf{x}-\gamma z)},\\
&\psi=\psi_0+(\delta \psi_1+i \delta \psi_2) e^{(i \textbf{k}_\perp\cdot \textbf{x}-\gamma z)},
\end{aligned}
\end{eqnarray}  

where the `background' envelope $\psi_0$ is real. These lead to a set of linear equations
\begin{eqnarray}
\begin{aligned}
&2 k_0\gamma \delta \psi_2 - k_\perp^2 \delta \psi_1-k_0^2 \frac{\delta n}{n_c}\psi_0=0,\\
&2 k_0\gamma \delta \psi_1+k_\perp^2 \delta \psi_2=0,\\
&n_0 \delta T+T_0 \delta n = -\varepsilon_0 \frac{n_0}{n_c}\psi_0 \delta\psi_1,\\
&\underline{\underline{\mathbf{\kappa}}}^0:\textbf{k}_\perp \textbf{k}_\perp \delta T=\varepsilon_0 \frac{n_0}{n_c}\nu_{ei}^0\psi_0\delta\psi_1.
\end{aligned}
\end{eqnarray}

These equations are then used to derive the filamentation dispersion relation with spatial growth rate $\gamma$,
\begin{equation}
4 k^2_0\gamma^2 =k_0^2 k_\perp^2 \left(\frac{n_0}{n_c}\right)^2\frac{\varepsilon_0 |\psi_0|^2}{n_0 T_0}\left(1+\frac{n_0\nu^0_{ei} }{\underline{\underline{\mathbf{\kappa}}}^0:\textbf{k}_\perp \textbf{k}_\perp}\right)-k_\perp^4.
\end{equation}
This expression can be simplified by defining the angle $\theta$ between the magnetic field direction and the perturbation vector $\textbf{k}_\perp$, this allows us to write
\begin{equation}
\underline{\underline{\mathbf{\kappa}}}^0:\textbf{k}_\perp \textbf{k}_\perp=k_\perp^2 \kappa_{clas}(\kappa^c_\parallel \cos^2 \theta+\kappa^c_\perp \sin^2 \theta).
\end{equation}
Considering now only the magnetic fields parallel with the laser wavevector (that is, when the angle $\theta = \pi/2$) the dispersion relation can be written
\begin{equation}
4 k^2_0\gamma^2 =k_0^2 k_\perp^2 \left(\frac{n_0}{n_c}\right)^2\frac{2 I_0}{c n_0 T_0}\left(1+\frac{1}{\lambda_{ei}^2 \kappa^{(1)}\kappa^c_\perp k_\perp^2} \right)-k_\perp^4.
\end{equation}
We have used the vacuum relation
\begin{eqnarray}
I=\frac{1}{2}c \varepsilon_0 |\psi|^2.
\end{eqnarray}
This dispersion relation includes both the ponderomotive and thermal mechanisms, with the latter modified by the normalised magnetised thermal conductivity $\kappa^c_\perp$. 

\begin{figure}
\includegraphics[width=\columnwidth]{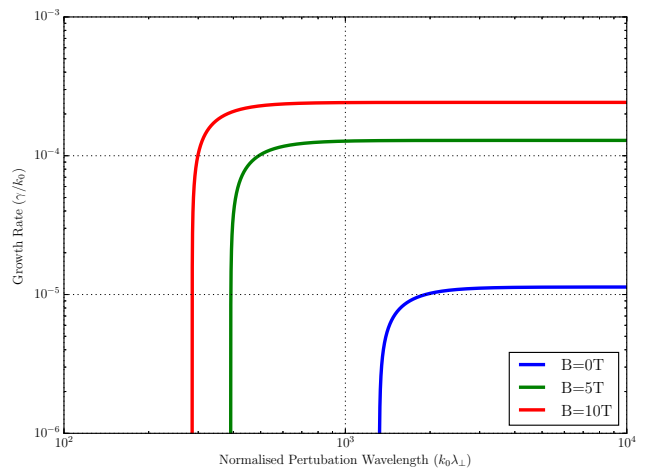}
\caption{ The filamentation growth rate of the thermal mechanism plotted against the perturbation wavelength shows that as the applied magnetic field increases, the peak growth rate increases and the cutoff is shifted to shorter wavelengths.}
\label{fig:grate}
\end{figure}

Fig.\ref{fig:grate} shows the spatial growth rate of a $1 \times 10^{14} \mathrm{W cm}^{-2}$ laser undergoing filamentation due to the thermal mechanism alone under a 0, 5 and 10T applied axial field. With an electron density of $1\times 10^{20} \mathrm{cm}^{-3}$ and a temperature of 1 keV this corresponds to Hall parameters of $\chi =$ 0, 11.9 and 23.9. It is plotted against the perturbation wavelength, defined as $\lambda_\perp=\frac{2\pi}{k_\perp}$, normalised with the laser wavenumber $k_0$. The thermal mechanism dispersion relation is obtained by ignoring the factor of unity in Eq. 32.

As the term $\kappa^c_\perp$ is reduced through magnetisation the maximum growth rate will increase and the growth rate will cutoff at higher perturbation wavenumbers. Indeed the maximum spatial growth rate under a 10 Tesla field is two orders of magnitude greater than the unmagnetised plasma.
In the highly magnetised case where approximately $\kappa^c_\perp\sim 1/\chi^2$ the maximum thermal growth rate is
\begin{eqnarray}
\gamma_{max}\approx\frac{n_0}{n_c}\sqrt[]{\frac{I_0}{2 c n_0 T_0}} \frac{\chi}{\sqrt{\kappa^{(1)}}\lambda_{ei}}.
\end{eqnarray}

\begin{figure}
\includegraphics[width=\columnwidth]{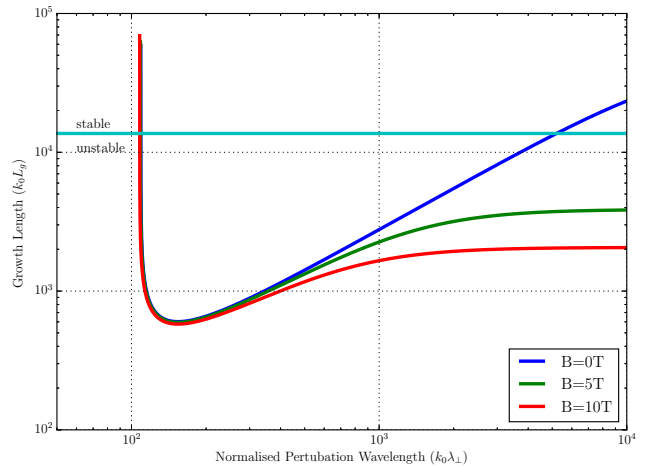}
\caption{ The dispersion relation shows the relationship between the growth length and perturbation wavelength for combined ponderomotive and thermal mechanisms. As the magnetic field is increased, the longer wavelength modes have a shorter growth length, meaning magnetised experiments are more susceptible to long perturbation wavelength filamentation.}
\label{fig:glength}
\end{figure}

It is instructive to look at the combined effect of both the ponderomotive and thermal mechanisms and their influence on the growth e-folding length, defined as
\begin{equation}
L_g=(2 \gamma)^{-1}.
\end{equation}
This parameter will allow us to compare the susceptibility of experimental parameters to this instability. If the scale length of the plasma is smaller than a couple of e-folding lengths, the experiment is susceptible to this instability. Fig. \ref{fig:glength} shows the growth length against perturbation wavelength for an applied axial magnetic field with a strength of 0, 5 and 10T. Superimposed onto the plot is a line representing 2mm. This has been chosen as  this is a characteristic experimental scale length in laser plasma experiments with magnetic fields \cite{Froula2007LaserPlasmas,Joglekar2016KineticHohlraums,Hohenberger2012}. Therefore the region of the plot above this line is stable to this instability since the length is too short for the instability to grow significantly over the length scale.

As can be seen in Fig. \ref{fig:glength} a significant proportion of the growth length curve dips into the unstable region. The effect of the magnetic field appears as a lowering of the instability growth length at longer perturbation wavelengths whereas the combined ponderomotive-thermal mechanism cutoff (where the combined mechanism growth rate equals zero) provides a lower limit to the influence of the magnetic field. This means that under a magnetic field the longer wavelength perturbation modes will grow faster and an experiment in which these modes are present will be more susceptible to filamentation. 

\subsection{Simulations of filamentation in a magnetised plasma}

To verify the linear analysis of this magnetised effect, the PARAMAGENT code was used to simulate a uniformly irradiated $Z=1$ plasma. A uniform 1 $\mu$m laser field of intensity $1\times10^{15}\mathrm{Wcm}^{-2}$ with small ($0.1\%$) harmonic perturbations heated the plasma with initial uniform electron density $1\times10^{20}\mathrm{cm}^{-3}$ under an magnetic field parallel to the laser wavevector. The initial temperature of the plasma was 20 eV and the magnetic field strengths used were 0, 5 and 10 Tesla.

\begin{figure}
\includegraphics[width=\columnwidth]{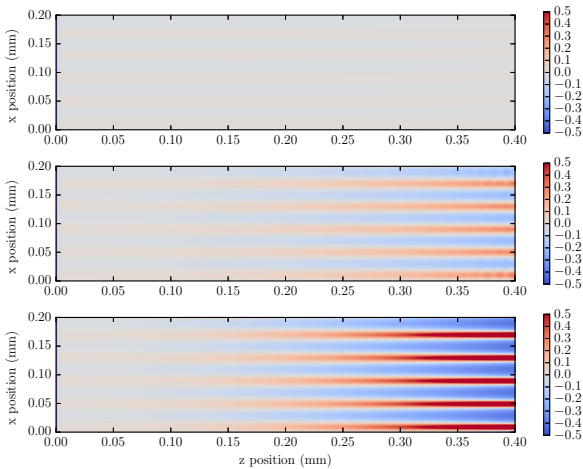}
\caption{ The filamentation of a uniform laser with $0.1\%$ perturbations for the unmagnetised (top), 5T (middle) and 10T (bottom) cases show the laser filaments with a shorter growth length as the magnetic field increases.}
\label{fig:filsim}
\end{figure}

Fig. \ref{fig:filsim} shows the output from the PARAMAGNET simulation after 100ps, It shows the deviation of the normalised laser intensity from the initial background under three different field strengths. The 10T field has caused the beam perturbations to grow to $50\%$ of the uniform background intensity over a length of only 0.4mm; whilst the unmagnetised case sees little variation over the same length.

\subsection{Kinetic Effects in Thermal Filamentation}

Kinetic effects on thermal transport are important in laser-produced plasmas\cite{Joglekar2016KineticHohlraums,Ridgers2008MagneticPlasmas,Brodrick2017TestingApplications}. Small scale structures in laser filamentation in the thermal regime mean the corresponding temperature profile will have characteristic length scales of the same magnitude as the electron-ion collision length. In unmagnetised plasmas this kinetic correction has been found to significantly alter filamentation \cite{Epperlein1990KineticPlasmas,Epperlein1991KineticPlasmas}. 

In this regime heat flow becomes nonlocal and the kinetic modification to the thermal conductivity must be included. It is therefore a useful question to ask what is the influence of nonlocal transport effects on the magnetised thermal mechanism filamentation.

The nonlocality of thermal transport in the absence of magnetic fields is encapsulated in the dimensionless nonlocality parameter $\sqrt{Z} k \lambda_{ei}$, defined as the product of the wavenumber of a perturbation in the plasma temperature $k$ and the electron delocalisation mean free path $\lambda_{ed}=\sqrt{\lambda_{ee}\lambda_{ei}}=\sqrt{Z}\lambda_{ei}$. For higher values of this parameter the thermal conductivity drops off relative to the Spitzer-Harm result according to an expression introduced by Epperlein \cite{Epperlein1992NonlocalPlasmas},
\begin{eqnarray}
\frac{\kappa_{nl}}{\kappa_{sh}} = \frac{1}{1+(30 \sqrt{Z}k \lambda_{ei} )^{4/3}}.
\end{eqnarray}

However, when a plasma is magnetised, the heat flow of the plasma is localised\cite{Schurtz2007RevisitingConditions,Kingham2002NonlocalGradients}. These two effects, the reduction of the heat flow via nonlocality and via magnetisation combine and the dual influence on filamentation must be considered. Kinetic, magnetised thermal transport has been considered by Brantov\cite{Brantov2003LinearPlasma} however only a simplified phenomenological form is required.

Therefore we propose a phenomenological function for the thermal conductivity perpendicular to the magnetic field relative to the local, unmagnetised conductivity
\begin{eqnarray}
\frac{\kappa_{\perp nl}}{\kappa_{clas}}=\frac{a}{1+b(\sqrt{Z}k_\perp \lambda_{ei})^c+d\chi^e}. 
\end{eqnarray}
This model includes the combined effect of the reduction from the magnetisation ($\chi$) and via nonlocality ($k_\perp \lambda_{ei}$).

In order to find the parameters $a,b,c,d,e$ the Epperlein-Short test\cite{Epperlein1992NonlocalPlasmas} is performed. The Epperlein-Short test aims to find the nonlocal thermal evolution of a 1D plasma using a numerical Vlaslov-Fokker-Planck (VFP) model. Given the requirements for a magnetised plasma, we use the IMPACT VFP code\cite{Kingham2004}. IMPACT solves the electron VFP equation in the diffusion approximation alongside the full maxwell's equations allowing the inclusion of magnetic field phenomena.

\begin{figure}
\includegraphics[width=\columnwidth]{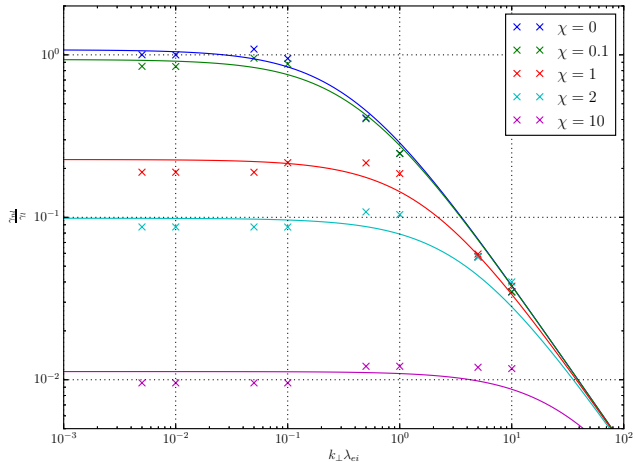}
\caption{The kinetic decay rate of a small temperature perturbation relative to the local analytic result against the nonlocality parameter. Data for IMPACT simulations with five different Hall parameters, 0, 0.1,1,2 and 10. The fit is overlaid on the data. The magnetisation localises and reduces the thermal conductivity, the nonlocal curl off of the conductivity is shifted down and to the right.}  
\label{fig:impactfit}
\end{figure}

We use Broderick's \cite{Brodrick2017TestingApplications} approach, in which a 1D hydrogen (Z=1) plasma is initialised with a small ($0.1\%$) sinusoidal temperature perturbation and is allowed to decay over multiple collision times. A static and uniform magnetic field of varying strengths is applied perpendicular to the temperature perturbation. The decay rate of the sinusoidal perturbation is proportional to the thermal conductivity of the plasma via
\begin{eqnarray}
\frac{\kappa_{\perp nl}}{\kappa_{clas}}=\frac{\gamma_{ nl}}{\gamma_{clas}}.
\end{eqnarray}
The local unmagnetised decay rate for a sinusoid with wavenumber $k$ is defined as
\begin{eqnarray}
\gamma_{clas}=\frac{2}{3}k_\perp^2 \frac{128}{3\pi}\zeta(Z) v_{th}\lambda_{ei}.
\end{eqnarray}
The nonlocal decay rate $\gamma_{nl}$ is found by fitting a decaying sinusoid to the output of the IMPACT simulations. The output from the IMPACT simulations are shown in Fig. \ref{fig:impactfit}. The function Eq. 38 is fitted to the data and the values are found to be $a= 1.116, b=2.73,c=1,d=3.72,e=1.4$. The result is overlaid on Fig. \ref{fig:impactfit}. The use of the diffusion approximation on the IMPACT code limits the accuracy of the decay in the high $k_\perp \lambda_{ei}$ regime, as such the data is fitted to runs ups to values of $k_\perp \lambda_{ei}=10$.

\begin{figure}
\includegraphics[width=\columnwidth]{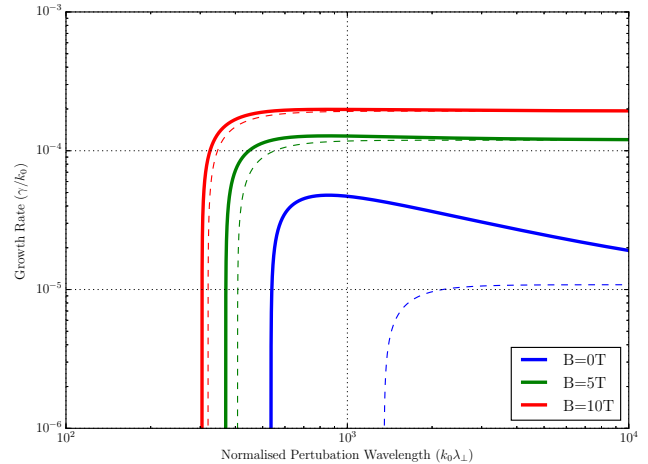}
\caption{The filamentation growth rate of the kinetic thermal mechanism (solid lines) plotted against the perturbation wavelength shows that in the unmagnetised case, the peak growth rate is greater relative to the local case (dashed lines) and the cutoff is shifted to shorter wavelengths. This nonlocal effect is suppressed in the magnetised curves.}
\label{fig:kgrate}
\end{figure}

The thermal filamentation dispersion relation using this nonlocal expression now has both a magnetisation and nonlocality parameter dependence and the growth rate is plotted in Fig. \ref{fig:kgrate}. In the local limit ($k \lambda_{ei}<<1$) the curves all plateau with a value determined by the magnetisation. As the nonlocality parameter increases the curve drops off, the point where the curve begins to drop off is shifted to the right at higher magnetisation values. This is the result of the magnetisation localising the thermal conductivity.

Looking now at the plot of the dispersion relation of thermal filamentation in isolation in Fig. \ref{fig:kgrate}, the most obvious difference as compared to the local thermal mechanism is the much higher peak growth rate. Also unlike the local case in Fig. \ref{fig:grate} where the curve flattened to a constant at long wavelengths, the kinetic growth rate converges towards the local rate at longer wavelengths. Whilst this is subdued in the magnetised curves, the influence is still noticeable. In the absence of a magnetic field the dispersion curve cutoff is shifted far to the left and has a higher peak growth rate.

\begin{figure}
\includegraphics[width=\columnwidth]{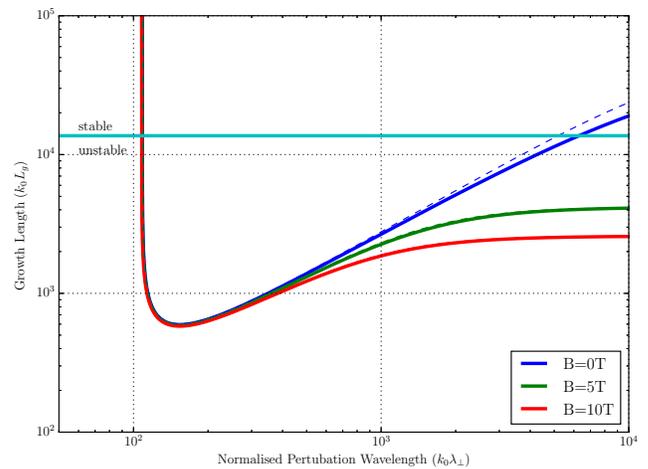}
\caption{The combined ponderomotive and thermal filamentation dispersion relation with the kinetic corrections (solid lines) shows some deviation from the classical case (dashed lines) in the longer wavelength region}
\label{fig:kglength}
\end{figure}

If we now compare the mixed ponderomotive and thermal growth length curves between Fig. \ref{fig:glength} and Fig. \ref{fig:kglength} the differences compared with the local case are most evident at longer perturbation wavelengths. The curve shows some deviation from the local curve in the long wavelength region whilst the low wavelength cutoff is unchanged.

\section{Magnetised self focusing in ICF experiments}\label{sec:influence}

In recent years the use of magnetic fields has been suggested as a means to improve the performance of inertial confinement fusion experiments. We will discuss the significance of the reported modifications to laser-propagation due to magnetised transport in the context of two schemes, the pre-pulse stage of the MAGLIF scheme\cite{Slutz2010a} and magnetised hohlraums\cite{Chang2011FusionImplosions}.  

The MAGLIF scheme necessarily has the laser pre-pulse propagating through a low Z plasma with a coaxial 10T magnetic field. The purpose is to pre-heat the fuel and as such uniform and deep penetration is required for effective heating. In such experiments the plasma is strongly magnetised and given the laser wavelength and intensity sits well within the unstable region of thermal regime filamentation. In the filament the local intensity can reach multiples of the initial intensity.

Taking as example the parameters of pre-heat experiments performed on the OMEGA-EP laser\cite{Harvey-Thompson2015a}, a $1.3\times10^{14}\mathrm{Wcm}^{-2}$ intensity laser propagating through a plasma with $n_e=2.5\times 10^{20}\mathrm{cm}^{-3} (0.025n_c)$ under an axial magnetic field of 7.5T. When a temperature of 400 eV is used, the maximum thermal growth length calculated using Eq. 34 is 1.01 mm, much smaller than the scale length of the 8mm fuel capsule used in the experiment.

Parametric instabilities such as Stimulated Raman Scattering (SRS) and Stimulated Brillouin Scattering (SBS) have an intensity threshold. The choice of laser intensity in the pre-heat stage is partially to undercut this threshold and mitigate the influence of SRS and SBS backscatter. This backscatter can reduce the absorption of the laser light into the MAGLIF fuel cell by scattering it out of the fuel and also reducing the depth of the penetration of the laser into the fuel cell. 

Therefore the magnetisation can lead to more backscatter as the local filament intensity exceeds the parametric instability threshold; as such the laser penetration will shorter and the absorption less effective compared to the unmagnetised case.

In a hohlraum environment the effect of magnetic fields can likewise influence laser propagation and absorption. In such experiments the magnetic field can be applied externally\cite{Chang2011FusionImplosions,Hohenberger2012} or self-generated \cite{Farmer2017SimulationEnvironment}; the orientation of the magnetic field will not be exactly parallel to the laser wavevector as investigated in this work, however the influence of the magnetised thermal conductivity will similarly drive down the instability threshold for thermal regime filamentation, attenuated by a trigonometric factor to account for orientation. Taking as example the parameters of the magnetised helium gas-fill hohlraum simulations performed by Strozzi \cite{Strozzi2015ImposedHohlraums}, the maximum thermal filamentation growth length is 0.82 mm, relative to the hohlraum length of 1 cm.

Furthermore, in direct drive experiments the laser uniformity is important for ensuring the reduction of laser-imprint seeded hydrodynamical instabilities\cite{Craxton2015}. The phenomenon of laser filamentation can cause small scale non-uniformity to grow in the underdense coronal plasma.The magnetisation increases the spatial growth rate of filamentation thus putting greater constraints of laser uniformity in the case of an experiment with an applied magnetic field.

\section{Conclusions}
In summary, we investigated how the effect of a magnetised thermal conductivity influences the propagation of a laser through an underdense plasma in regimes common in magnetised high energy density physics experiments. By using an analytic model of the self-focal point of a Gaussian beam derived from a steady-state fluid-plasma paraxial-laser model, the focal point is shown to shorten by a factor proportional to the square root of the perpendicular thermal conductivity of the magnetised plasma; this is a direct result of the anisotropic magnetised transport of Braginskii.  This in turn means for a highly magnetised plasma, the focal length relative to the unmagnetised case is approximately inversely proportional to the magnetic field strength parallel to the wavevector of the laser. 

Using the PARAMAGNET laser-plasma transport code, simulations of this self focusing effect were performed with a Gaussian laser in a plasma with a range of magnetic field strengths. In these simulations, the self-focal point from the analytic model is retrieved whilst showing nonlinear behavior such as repeated defocusing and refocusing not present in the analytic model. The quantitative difference in focusing behaviour resulting from magnetisation of the thermal conductivity is manifest. 

Similarly a linear model of thermal filamentation of a laser was derived and the dispersion relation follows the same asymptotic dependence of the normalised thermal conductivity leading to an increase in the thermal mechanism growth rate as electrons are increasingly magnetised. When combined with the ponderomotive mechanism the effect of magnetisation means the filamentation of long-wavelength perturbations is particularly significant. Simulations performed with the PARAMAGNET code of the linear regime of filamentation of a plasma heated by a uniform laser under a magnetic field with harmonic perturbations yields an order of magnitude shortening of the e-folding length for even low values of the Hall parameter in the range of 0.1-1.

Considering the nonlocality of thermal transport in laser-plasma interactions in the regimes typical in the experiments noted above, the Epperlein-Short test was performed to find a phenomenological fit for the perpendicular thermal conductivity in plasma where both kinetic and magnetic effects are included. When this fit is used in the filamentation dispersion relation it shows even shorter e-folding lengths in the long-wavelength limit compared to the local dispersion relation.

When considering these effects in the context of magnetised laser fusion experiments  it is found they sit in regimes susceptible to significant laser focusing exacerbated by magnetisation. The resulting filamentation produces localised  intensities much higher than the threshold for parametric instabilities. These instabilities are detrimental to laser absorption and penetration by the laser by backscattering the laser light out of the fuel or Hohlraum. As such it is important in the modeling of high-energy-density laser plasma experiments to include the influence of full thermal magnetised transport not just on the plasma but also on the laser propagation.

\begin{acknowledgments}
The authors thank D. Hill and Dr M. Read for helpful comments on this work. This work was funded through ESPRC doctoral training grant (No. EP/M507878/1) and via an AWE plc CASE partnership. Computing resources were provided by the Imperial College Research Computing Service, DOI:10.14469/hpc/2232
\end{acknowledgments}
\appendix
\section{Derivation of the normalised Gaussian beam waist}
After substituting the linearised momentum equation into the linearised energy equation, using the form of $\psi$ defined in Eq. 12, the equation for the density change is thus
\begin{equation}
\frac{1}{r}\frac{\partial}{\partial r}\left(r \frac{\partial \delta n}{\partial r}\right)=\frac{1}{2}\frac{n_0 }{T_0 n_c}\varepsilon_0|\psi|^2\left(\frac{n_0\nu^0_{ei}}{\kappa_0}+\frac{4}{ a^2_0 \alpha^2}\left(1-\frac{r^2}{a^2_0 \alpha^2}\right)\right).
\end{equation}
In order to solve this we use the close-to-axis assumption
\begin{equation}
|\psi|^2\approx\frac{A^2_0}{\alpha^2}\left(1-\frac{r^2}{a^2_0 \alpha^2}\right).
\end{equation}
The second order dielectric variation required to find the equation for the beam variance  is then
\begin{equation}
\delta\epsilon_2 = -\frac{1}{8}\left(\frac{n_0}{n_c}\right)^2 \frac{\varepsilon_0 A_0^2}{n_0 T_0 \alpha^2}\left(\frac{n_0 \nu^0_{ei}}{\kappa^0}+\frac{4}{a_0^2 \alpha^2}\right).
\end{equation}
This is then used with Eqs. 22 and 24 to get 
\begin{equation}
\alpha''=\frac{c_1}{\alpha^3}-\frac{c_2}{\alpha},
\end{equation}
with
\begin{eqnarray}
\begin{aligned}
&c_1=\frac{1}{k^2_0a^4_0}-\frac{1}{2}\left(\frac{n_0}{n_c}\right)^2\frac{\varepsilon_0 A^2_0}{ T_0 a^2_0},\\
&c_2=\frac{1}{8}\left(\frac{n_0}{n_c}\right)^2\frac{\varepsilon_0 A^2_0}{ n_0 T_0}\frac{n_0 \nu^0_{ei}}{\kappa_0}.
\end{aligned}
\end{eqnarray}
This can be written
\begin{equation}
\left(\frac{d \alpha^2}{d z}\right)^2=- c_2 \alpha^2 \ln\alpha^2 +c_3 \alpha^2 -c_1.
\end{equation}
In order to find an analytic result, the approximation to the natural log about $\alpha =1$ is used
\begin{eqnarray}
\alpha^2 \ln \alpha^2 \approx \alpha^2(\alpha^2-1),
\end{eqnarray}
which yields
\begin{equation}
\left(\frac{d \alpha^2}{d z}\right)^2=- c_2\alpha^4+ (c_3+c_2) \alpha^2 -c_1,
\end{equation}
and the solution is found to be
\begin{equation}
\alpha^2=d_1+d_2\sin(d_3 z + d_4),
\end{equation}
with constants $d_1-d_4$ defined as
\begin{eqnarray}
\begin{aligned}
&d_1=\frac{\frac{1}{R^2}+c_1+c_2}{2 c_2},\\
&d_2=\sqrt{d_1^2-\frac{c_1}{c_2}},\\
&d_3=\sqrt{c_2},\\
&d_4=\arcsin(\frac{1-d_1}{d_2}).
\end{aligned}
\end{eqnarray}

In the case of a long geometric focus such that the Rayleigh length is very long relative to the propagation distance, $R\rightarrow\infty$. These expressions can be reduced to
\begin{eqnarray}
\begin{aligned}
&d_1=\frac{1}{2}(1+\frac{c_1}{c_2}),\\
&d_2=\frac{1}{2}(1-\frac{c_1}{c_2}),\\
&d_3=\sqrt{c_2},\\
&d_4=\frac{\pi}{2}.
\end{aligned}
\end{eqnarray}
\bibliography{Mendeley}

\end{document}